# Frequency-specific, valveless flow control in insect-mimetic microfluidic devices

Krishnashis Chatterjee[1,3], Philip M. Graybill [2,4], John J. Socha[3], Rafael V. Davalos[2,3] and Anne E. Staples[1,3]

[1]Laboratory for Fluid Dynamics in Nature, Biomedical Engineering and Mechanics, Virginia Tech, Blacksburg, VA, USA.

[2]Bioelectromechanical Systems Laboratory, Biomedical Engineering and Mechanics, Virginia Tech, Blacksburg, VA, USA.

[3]Biomedical Engineering and Mechanics, Virginia Tech, Blacksburg, VA, USA.

[4]Mechanical Engineering, Virginia Tech, Blacksburg, VA, USA.

E-mail: aestaples@vt.edu

**Abstract**

Inexpensive, portable lab-on-a-chip devices would revolutionize fields like environmental monitoring and global health, but current microfluidic chips are tethered to extensive off-chip hardware. Insects, however, are self-contained and expertly manipulate fluids at the microscale using largely unexplored methods. We fabricated a series of microfluidic devices that mimic key features of insect respiratory kinematics observed by synchrotron-radiation imaging, including the collapse of portions of multiple respiratory tracts in response to a single fluctuating pressure signal. In one single-channel device, the flow rate and direction could be controlled by the actuation frequency alone, without the use of internal valves. Additionally, we fabricated multichannel chips whose individual channels responded selectively (on with a variable, frequency-dependent flow rate, or off) to a single, global actuation frequency. Our results demonstrate that insect-mimetic designs have the potential to drastically reduce the actuation overhead for microfluidic chips, and that insect respiratory systems may share features with impedance-mismatch pumps.

## 1. Introduction

Microfluidic technology is expected to play a critical role in the cooling of integrated circuits, allowing Moore's law to persist past 2021, [1,2] and in other vitally important applications such as lab-on-a-chip interventions in global health and environmental monitoring [3–9] fundamental topics in microfluidics, such as efficient strategies for mixing and flow control at the microscale, are still current areas of investigation and their solutions are necessary for achieving such applications. One issue is that microfluidic technology suffers from an actuation overhead problem in which microfluidic chips are tethered to extensive off-chip hardware. Such hardware incurs monetary costs and requires physical space, which can be especially problematic for lab-on-a-chip systems where space is a premium, for example in scientific payload for planetary probes [61, 62]. State-of-the-art microfluidic large-scale integration (mLSI) and microfluidic very-large-scale integration (mVLSI) chips contain thousands of flow channels that each require three separate actuations to control the rate and direction of the flow within them [10,11].

Significant progress has been made both in scaling up microfluidic chips to vLSI dimensions, and in reducing the amount of peripheral actuation machinery associated with microfluidic devices. The largest mVLSI chips are now millions of valves per square centimeter [11–13]. Actuation strategies have been designed that reduce the required actuation load from three actuations per flow channel to one actuation per chip, when combined with check valves [14–16], enabling advances in pneumatically actuated, passive elastomeric microfluidic devices for mLSI and vLSI chips.

In contrast to these engineering efforts, insects can be viewed as nature's testbed for the active handling of fluids at the microscale. The honeybee, as an example, expertly manipulates air, water, nectar, honey, wax, and hemolymph at the microscale. Insect flight is the most demanding activity known, and the aerobic scope of insects is unrivaled in the animal kingdom [17]. The ratio of maximum to basal rate of respiration in many species of locusts, bees, and flies is in the range 70–100 [17,18] whereas in humans this ratio approaches 20 maximally, and other small mammals and birds attain only about a 7- to 14-fold increase in metabolic rate during maximum exertion [17,19]. Among many reasons for their superior performance, such as effective coupling of adenosine triphosphate (ATP) hydrolysis and regeneration in the working flight muscles [17], insects generally do not use blood as an intermediate oxygen carrier [20]. Instead, they transport freshly oxygenated air from a series of spiracular openings directly to the tissues through a complex network of thousands of respiratory tracts called tracheae, which ramify and decrease in size as they approach the cells [21].

Although microfluidic device flow channel densities have approached those of insects, actuation efficiency and device performance lag far behind. Here, we sought to benefit from evolutionary advances made by insects in handling fluids at the microscale by incorporating some of the fundamental features of their unique respiratory systems into the design of a series of biomimetic microfluidic devices.

Additionally, our devices serve as microfluidic models that can provide new insight into the mechanisms that insects employ to control airflows in the tracheal system. With more than one million described species, insects represent the most diverse group of animals on earth [50]. Correspondingly, their respiratory systems exhibit a diverse array of morphologies and kinematics used for transport of gases to and from the tissues. Their tracheal systems comprise thousands of short sections of tracheal tubes and junctions that connect the ramifying and anastomosing network.

Species that employ rhythmic tracheal compression to produce advective flows are able to modulate both actuation frequency and degree of collapse in the system [22,24,29,51–54]. Furthermore, some species can produce one-directional flows through the network [55], whereby ambient air enters the tracheal system through one spiracle and exits the body through a different spiracle, a physiologically effective mechanism of gas exchange [56]. The discovery that flow direction can be controlled by frequency in a model tracheal network suggests a new hypothesis for flow control in the insect respiratory system.

Although visualizing tracheal wall displacement is possible using synchrotron x-rays, visualizing airflow patterns within these small channels in the insect has so far proven to be intractable [52]. Microfluidic models such as the ones presented here therefore provide a powerful new tool for studying advective flow production in insects, akin to the recent microfluidic platforms used to understand alveolar dynamics in human systems [57,58]. In turn, these models can lead to new principles of device design, demonstrating that insect-inspired microfluidics can provide a platform for 'mining' the biodiversity of transport solutions provided by evolution [59].

## 2. Background and Results

*2.1 Single-channel devices*

Inspired by insect respiratory mechanics (see Figure 1), we designed, fabricated, and tested a total of eleven single-channel devices (as shown in Figure 2(a)) using current state-of-the-art multilayer soft lithography techniques (see Figure 2(c) and supplementary materials). The positive flow is in the "+" direction. Devices S2 and S4-9 incorporate tapered flow channels to reproduce directional collapse, as observed in some insects. Devices S1 and S3–11 reproduce the discrete collapse phenomenon by incorporating two discrete collapse locations. Devices S1 and S6–11 incorporate a u-shaped actuation channel in order to produce a time lag between collapses. The specific geometries and representative dimensions of the eleven devices are provided in Supplementary Table 2 and Supplementary Figure 6, respectively.

These single-channel devices were meant to capture the fundamental kinematic and actuation strategies occurring in a single insect tracheal pathway. Tracheal collapse, while generally pathological in vertebrates, occurs during a cyclical form of active respiration known as rhythmic tracheal compression (RTC), found in some insects [22]. The collapse is hypothesized to occur in response to the rhythmic abdominal contractions that pressurize the hemolymph in the animal's body cavity, which surrounds the tracheae and causes them to buckle in localized regions [23–27] (see also Figures 1(a) and 1(c)–(e)). Within a body segment, the hemolymph pressure is a single scalar actuation input that appears to largely control the complex, passive dynamics of the respiratory network [28].

We were motivated by this efficient method of fluid handling and hoped to fabricate microfluidic devices that could simplify complex flow actuation methods at the microscale. To do this, we extended current three-layer PDMS technology by connecting the overlying actuation channels in the top layer to a single, global actuation chamber, so that they are all actuated simultaneously by the same source, at an actuation frequency $f$ and a differential pressure across the elastomeric membrane, $\Delta p$ (as shown in Figure 2(b)).

In addition to using a single pressurized actuation chamber, we incorporated both the directional and discrete collapse phenomena that have been observed [23–26,29] and modeled [30–33] in insects. Directional collapse (Figure 1(c)) is hypothesized to occur because of either a variation in material or structural properties along the axis of the respiratory tract, or because of pressure waves propagating through the hemolymph, or a combination of both phenomena. Here, we added directionality to the collapse of the channel's ceiling in some of the devices by fabricating tapered flow channels (devices S2 and S4-9, shown in Figure (2a)). To produce discrete collapses (Figure 1(d)–(e)), we fabricated devices with two separate sections of elastomeric membrane (devices S1 and S3–11, Figure 2(a)).

Some of the discrete collapse devices (S1 and S6–11, Figure 2(a)) exhibited a time lag between the occurrence of the first and second collapses in a contraction cycle. This time lag was accomplished by incorporating u-shaped actuation channels in these devices so that the pressurized gas (air or nitrogen) in the actuation channel would reach one collapse site slightly before the other, owing to the finite time required for the pressure wave to propagate through the gas. Additionally, there is an inherent lag in the timing of the membrane collapses in devices S3–S11 (Figure 2(a)) resulting from the different response times of the elastomeric membranes of different size. We estimated this difference to be $\tau_{large}/\tau_{small} \sim 25$ in the devices by approximating the deflecting portions of the membranes as rectangular in shape [34]. This estimate was confirmed from imaging in our experiments. The maximum deflections of the larger and the smaller membranes for different actuation pressures are plotted in Supplementary Figure 7. The lighter areas in the image series in Figure 3(d) show the collapsed regions of the ceiling of the microfluidic channel (made up of a thin PDMS membrane), while the darker areas are the uncollapsed regions. We observed that, for the larger collapse sites in the tapered-channel devices, the membrane collapsed at the ends of the membrane first, and the collapse then propagated inward toward the membrane's center during the collapse part of the cycle (Figure 3(d), $t$ = 0–0.022 s). During the re-expansion part of the cycle, however, the membrane re-expanded uniformly (Figure 3(d), $t$ = 0.028–0.072 s).

All eleven single-channel devices acted as pumps, producing a unidirectional flow. For a given actuation pressure (both in magnitude and duty cycle, held constant at 0.50 for all experiments), the flow rate in the devices depended on actuation frequency alone (Figure 2(b)).

In one device (S4, shown in Figure 2(a)), we were also able to control the flow direction solely by actuation frequency. At an actuation pressure of 10.0 ± 1.0 psi and actuation frequencies below a critical actuation frequency of about 4 Hz, device S4 produced forward flow. However, for actuation frequencies above that critical frequency, it produced flow in the reverse direction (Figures 3(a) and 2(e)), thereby acting as a valveless, reversible microfluidic pump.

In one device (S11, as shown in Figure 2(a)), we held the actuation frequency and duty cycle constant and varied the actuation pressure, and found that the flow rate could be controlled continuously with actuation pressure (Figure 3(c)).

## 2.2 Multichannel devices

Four multichannel devices were designed and fabricated (Figure 4) inspired by the basic geometric structure of the main thoracic tracheal network found in some beetles (here we use *Platynus decentis*, shown in Figure 1(a)), whose specific flow channels were designed after analyzing the results of the single-channel device flow rate experiments. Our aim was to switch flow off and on in an individual branch of the network by varying the global actuation frequency alone.

To accomplish this frequency-based flow switching, the designs of many of the individual channels in the network devices were based on the single-channel device S4, the reversible pump. Specifically, the parent channels in devices M2-M4 (paired channels labeled "C" in Figure 4(b)-(d)) used the design of device S9, oriented to provide positive flow for all global actuation frequencies (see the caption of Figure 4 for flow direction convention). The daughter channels (pairs labeled "A" and "B" in Figure 4(b)-(d)) in devices M2-M4 used the design of the single-channel device S4. In device M2, the orientation of the device S4 design in the inner daughter channels ("B") mirrored the orientation of the device S4 design in the outer daughter channels ("A"). The reasoning for this design choice was that the inner channels would presumably attempt to pump in the negative flow direction at low global actuation frequencies, but meet the positive flow from the parent channels ("C"), and the net result would be no flow through channels "B." Then, at frequencies greater than the critical reversal frequency of the "B" channels, positive flow would develop in channel "B." Similar considerations were made for the designs of devices M3 and M4.

In order to double the number of experiments that could be performed, experiments were performed in only half of the left-right symmetric devices. The devices were divided by etching away the connecting bridge between the two channels labeled "B" in Figure 4, resulting in half devices (Figure (5)). These network devices were then subjected to the same single (global) square-wave periodic actuation pressure signal as were the single channel devices, and the resulting flow rates in the constitutive channels were measured.

Two of the multichannel designs (M1 and M3) produced flow in a single branch of the network only above a critical frequency, demonstrating valveless, frequency-based flow switching. Below the critical frequency (around 0.7 Hz for device M1 and 0.5 Hz for device M3), there was no measurable flow through the channel in question (channel C for device M1 and channel B for device M3). The presence or absence of flow through channel B in device M3 is demonstrated in Figure 5(a)-(b). In Figure 5(a), the actuation frequency is above the critical value, and black fluid can clearly be seen passing from channel B into channel A. In Figure 5(b), the actuation frequency is below the critical value, and only a small amount of the black dye from channel B is seen to pass from the end of channel B, staying localized at the channel's exit. This small amount of leakage likely occurred due to axial diffusion, which is enhanced in oscillating flows even when no net mean flow is present [35].

## 3. Discussion

We fabricated a series of 11 single-channel and four multichannel microfluidic devices that mimicked key features of active ventilations in the insect respiratory system. Each of the single-channel devices acted as a microfluidic pump, transforming a symmetric actuation pressure signal in the control chamber fluid into a unidirectional flow in the fluidic channel. Importantly, this transduction resulted from asymmetry in the network geometry, either in the flow channel itself, or in the part of the actuation chamber in contact with the flow channel.

Furthermore, one of the single-channel devices (device S4) exhibited flow reversal above a critical actuation frequency, demonstrating the discovery of a pneumatically actuated, valveless, reversible, microfluidic pump. We hypothesize that the flow direction is determined by two factors: 1) the relative balance of upstream versus downstream hydraulic resistance, as the momentum is injected via the motion of the channel's ceiling roughly at the mid-channel location; and 2) the effect of nonlinear resonant wave interactions, as are found in impedance-mismatch pumping. The kinematic asymmetry observed in the collapse and re-expansion of the elastomeric membrane indicates that the hydraulic resistance in the channel will also exhibit temporal and spatial asymmetry during the collapse and re-expansion parts of the actuation cycle.

This reversible flow direction channel design was incorporated into the multichannel devices, resulting in four microfluidic chips that can be operated by frequency control, with the individual channels responding selectively to the single global actuation frequency. In two of these chips, a single channel can be switched on or off via a critical global actuation frequency.

The flow rates produced by our top-performing single channel device (S11) and our top-performing multichannel device (M2) compare favorably to historic and current state-of-the-art pneumatically actuated PDMS micropumps (Table 1). The normalized flow rates that we report here are obtained by dividing the flow rate in $\mu Lmin^{-1}$ by the cross-sectional area of the flow channel, the actuation pressure, and the actuation frequency, in order to make a fair comparison across devices and experimental conditions. As evidenced by this comparison, the normalized flow rate in our best-performing device (M2) is higher than all but two of the flow rates reported for the similar PDMS micropumps surveyed here.

Our results demonstrate a fundamentally different way to pump fluids at the microscale, using a simplified actuation scheme. Multilayer, pneumatically actuated microscale peristaltic pumps built using soft lithography were pioneered by Unger et al. [37]. Although there have been several innovations since then to reduce the number of needed controllers (e.g., [12,14,15]) generally these peristaltic pumps require three overlying pneumatic control channels for fine control of flow rate and direction in a single flow channel. Shortly after the first introduction of such pumps, the same group pioneered microfluidic large-scale integration (mLSI) in which multiplexors that work as a binary tree allow control of $n$ fluid channels with only $2 \log_2 n$ control channels, a type of quasi-static control [10]. In [10], control of a channel or chamber indicated that access to the chamber can be switched on or off, accomplished for $n$ chambers using $2 \log_2 n$ control channels. In later studies (e.g., [45]), such control was demonstrated to be possible using just $n$ control lines for $n!/(n/2!)^2$ flow chambers. In contrast to these devices, the devices presented here have precisely controlled flow rates and flow directions within multiple individual channels using a single control line, a fundamentally different type of dynamic control. Our results also suggest that, in principle, flow rate and direction of flow in an arbitrarily large number of fluid channels can be controlled with a single control chamber, leading to an $n^0$ rule.

Peristaltic pumps, such as those by Unger *et al*. and Thorsen *et al*. discussed previously, send a traveling collapse wave along the axis of the flow channel. The flow is always in the same direction as the traveling wave, and the average flow speed is equivalent to the wave speed [46]. Additionally, there is a linear relationship between the frequency of the compression wave and the flow rate it produces [47]. Our micropumps clearly violate these precepts: the wave of ceiling motion producing the flow cannot be described as a travelling wave, because the waveform changes as it propagates along the channel. (See Figure 3(d), which depicts one full cycle of collapse in device S4, clearly demonstrating a heterogeneous waveform.) In addition, the flow can reverse direction and is not always in the same direction as the collapse wave, and the flow rate produced is a nonlinear function of the actuation frequency. Rather than peristaltic pumping, our devices appear to share many features with impedance mismatch pumping, where nonlinear resonant wave interactions drive the flow, and the flow rate produced has a nonlinear relationship to the actuation frequency [47–49]. In device S4, there are different portions of the flow channel with different material properties and hence impedances, as occurs in impedance mismatch pumping [48]. The waves generated by the actuation of the thin membrane travel along the membrane, and after encountering the stiffer ends (with different impedance), get reflected back. The encounter between the travelling wave and the reflecting wave results in a pressure build up, which drives a flow.

It remains to be seen how far these results will scale up toward the full vLSI scale, with a single actuation providing rich, passive control of thousands of flow channels as in insect respiratory systems. Given the many differences between insects' complex three-dimensional respiratory morphology and the planar geometries of current vLSI microfluidic devices, this may require creative modeling efforts. Regardless, insect-inspired control strategies may provide a key to developing microfluidic platforms that carry out heterogeneous fluidic operations in response to a single, global actuation input. For example, such strategies may lead to the development of platforms for carrying out multiple genomic and proteomic analyses in parallel using a single fluid sample, and with a very low actuation cost. These smart, bioinspired control strategies could also lead to the first truly portable, self-contained labs-on-a-chip, providing insect-style control in insect-sized packages.

Despite the challenges of transferring complex insect microfluidic control strategies to engineered devices, many more fundamental aspects of insect respiratory systems remain ripe for investigation and application in gaseous microfluidics, including the role of the small (~1 µm diameter) but numerous tracheoles, and uneven wall features (e.g., helical or circumferential windings called taenidia), which may contribute to mixing, heat, and mass transfer.

The results presented here suggest that we should continue to look to insect respiratory mechanics for clues about efficient geometries and strategies when scaling microfluidics up to three dimensions, advancing a broad range of critical microfluidics applications, such as integrated circuit cooling.

## 3. Materials and Methods

### 3.1 Microfluidic devices

Standard photolithography and microfabrication techniques were used to fabricate the PDMS-based microfluidic devices used in the experiments [60]. Negatively patterned master molds for the actuation and insect-network (fluidic) channels were created using photolithography by spinning SU-8 2035 (MicroChem) on silicon wafers to create a pattern approximately 80 $\mu$m in depth. Polydimethylsiloxane (PDMS) (Dow Corning, Sylgard 184) was mixed in a 10:1 weight ratio of base to cross-linker. Afterwards, it was cast-molded on the silanized master molds, cured, and slowly peeled off. The inlet and the outlet holes of the microfluidic devices were punched using a 0.75 mm biopsy punch. To create the thin PDMS layer (approximately 14–20 $\mu$m thick) between the actuation and fluidic channels (Figure 2(b)), a silanized silicon wafer was spin-coated with PDMS, mixed in 5:1 weight ratio of base to cross-linker at 3000 rpm for 60 seconds. The PDMS was then cured on a hot plate at 90°C for 30 minutes. The actuating channel was bonded to the middle membrane using a plasma cleaner (PDC-001, Harrick Plasma). The actuating channel and membrane assembly was then again plasma-bonded to the fluidic channel after carefully aligning them to their desired positions, after which the entire device was bonded to a glass slide (Figure 1(b)). The microfluidic devices were kept in vacuum prior to experiments.

*3.2 Experimental setup*

The actuating channels were pressurized using nitrogen gas and depressurized by vacuum through a single port, which served both as an inlet and an outlet. The pressure of the nitrogen was regulated via a precision regulator (McMaster Carr, 2227T21). The pressure and vacuum range of the pressure gauge/regulator (as mentioned in the manual) is -30 to 30 psi. In order to switch between positive and vacuum gauge pressure, the actuation channel was connected via tubing (Cole Parmer, AWG 30) to a fast-acting solenoid valve (FESTO, MHE2-MSIH-5/2-QS-4-K 525119 D002). A 24 V power supply was used to power the solenoid valve, which was computer-controlled by a microcontroller using a solid-state relay (Arduino, Board model: UNO R3).

*3.3 Experimental Method*

Before conducting each run, the devices were primed using ethanol to remove bubbles. Food coloring mixed with water was used as the working fluid in the fluidic channels. The inlet and the outlet ports of the fluid channels were connected to short tubes. The flow rate, produced by the actuation of the thin membrane in localized areas on top of the fluid channel, was determined by measuring the displacement of the fluid front in the outlet tube over a fixed amount of time. Displacement was measured by placing the outlet tube parallel to a measuring ruler with graduations. At least three readings were taken for a single data point, and the average of these readings was used to calculate the flow rate. Out of twelve S4 devices that were fabricated and tested, seven devices demonstrated frequency-dependent flow reversal. The five devices that did not demonstrate flow reversal can be attributed to some misalignment in the positioning of the actuation channels on top of the fluidic channels, which had to be done manually within two minutes after taking the components out of the plasma cleaner. The devices were also tested for repeatability. The flow rate data were taken while the frequencies and the pressures were varied from low to high and then again from high to low. The devices showed the same pattern of behavior in both cases. High resolution video of the device performance was captured using an Edgertronic High Speed Video Camera (Sanstreak Corporation) and Nikon lens (AF MICRO NIKKOR, 60 mm, 1:2.8 D).

**Author contributions**

A.E.S., J.J.S., R.V.D., and K.C. designed the project; K.C. and P.M.G. designed the experiments; K.C., P.M.G, R.V.D., and A.E.S. designed the microfluidic devices; P.M.G. and K.C. fabricated the microfluidic devices; K.C. performed the flow rate experiments. K.C. carried out the flow visualizations with the help of J.J.S.; K.C. and A.E.S. analyzed the resulting data; A.E.S., K. C., and P.M.G. wrote the manuscript; K.C., J.J.S., R.V.D., and P.M.G. revised and edited the manuscript; A.E.S., J.J.S., and R.V.D. oversaw the project; A.E.S. supervised the investigations of fluid mechanics, R.V.D supervised the microfluidic device studies, and J.J.S. supervised the biological content of the study. **Conflict of interest:** The authors declare that they have no conflict of interest. **Data and materials availability:** All data are available in the main text or the supplementary materials.

**Acknowledgements**

**General:** The authors thank Mohammad Bonakdar and Joel Garrett for useful discussions and research support, and Kate Lusco for designing Fig. 2c. **Funding:** This work was supported by the US National Science Foundation's Chemical, Bioengineering, Environmental and Transport Systems Division (1437387), Emerging Frontiers in Research and Innovation program (0938047), and Integrative Organismal Systems program (1558052).

**Tables and Figures**

Table 1 Comparison of maximum normalized flow rate among state-of-the-art pneumatically actuated microfluidic pumps. Entry number (N) corresponds to flow rate ranking, with N = 1 indicating the lowest flow rate and N = 11 the highest.

| N | Source (Author, journal, year) | Device description | Normalized flow rate ($\mu L/min)/\mu m^2/Hz/psi$ |
|---|---|---|---|
| 1 | Lee et. al, Lab Chip 18, 2018 [36] | 3D printed Quake style valve | $9.3 \times 10^{-8}$ |
| 2 | Unger et al., Science 288, 2000 [37] | Elastomeric peristaltic micropump | $1.0 \times 10^{-7}$ |
| 3 | Jeong & Konishi, Micromech. Microeng. 18, 2008 [38] | Peristaltic micropump, actuation regions separated by serpentine channels | $4.1 \times 10^{-7}$ |
| 4 | Chiou et al., Micromachines 6, 2015 [39] | Double-side mode PDMS micropump | $5.3 \times 10^{-7}$ |
| 5 | Wang & Lee, Micromech. Microeng. 16, 2006 [40] | Pneumatically driven peristaltic micropump with serpentine actuation channels | $1.5 \times 10^{-6}$ |
| 6 | Lai & Folch, Lab Chip 11, 2010 [41] | Single-stroke peristaltic PDMS micropumps | $3.2 \times 10^{-6}$ |
| 7 | Present work (device S11) |  | $3.7 \times 10^{-6}$ |
| 8 | Huang et al., Micromech. Microeng.18, 2008 [42] | Membrane-based serpentine shaped pneumatic micropump | $7.1 \times 10^{-6}$ |
| 9 | Present work (device M2) |  | $7.1 \times 10^{-6}$ |
| 10 | Huang et al., Micromech. Microeng. 16, 2006 [43] | Pneumatic micropump with serially connected actuation chambers | $5.7 \times 10^{-6}$ |
| 11 | So et al., Lab Chip 14, 2014 [44] | Caterpillar locomotion-inspired valveless micropump, teardrop-shaped elastomeric membrane | $2.5 \times 10^{-5}$ |

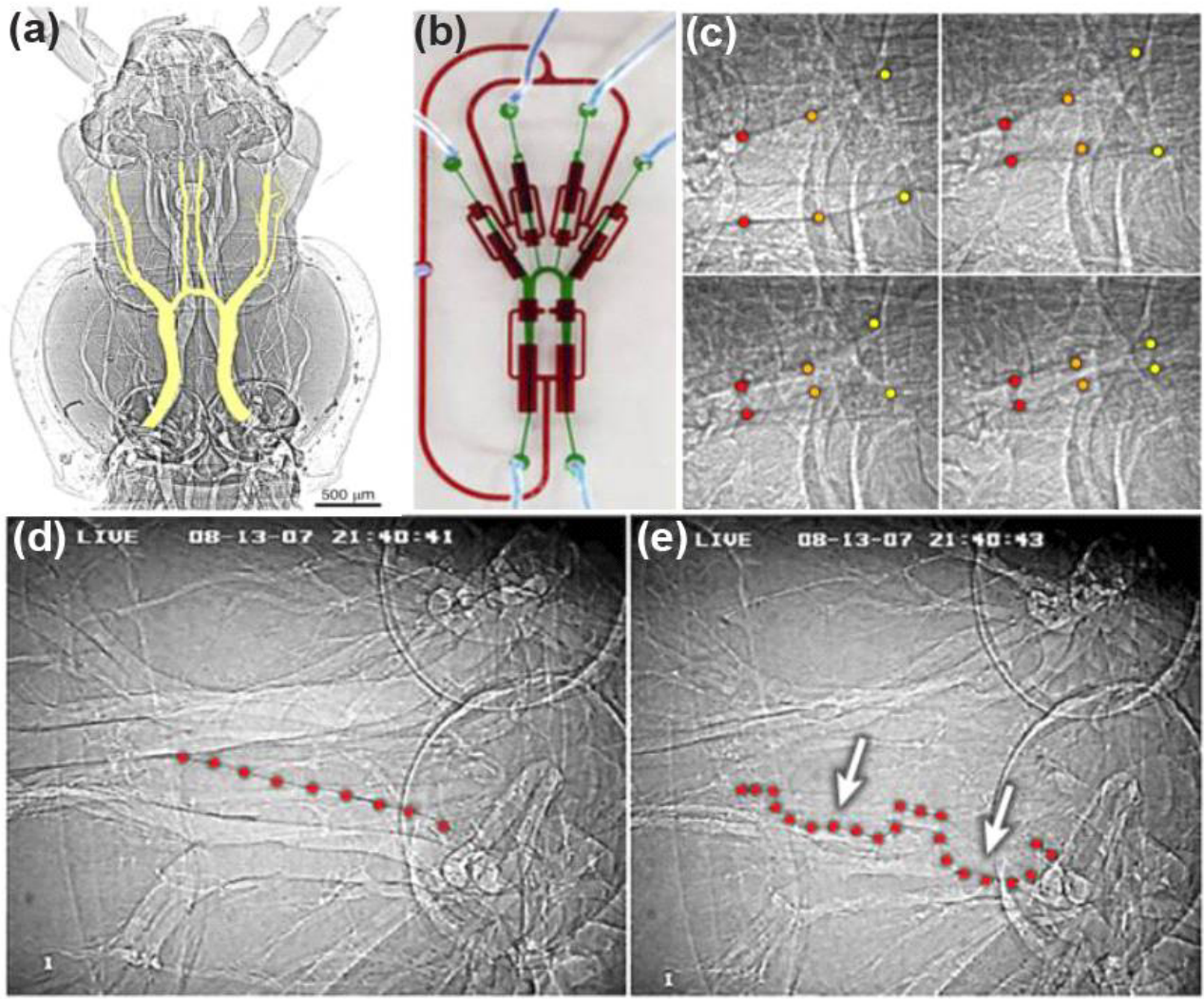

**Figure 1 Tracheal collapse in insects and design of insect inspired microfluidic devices.** (**a**) Synchrotron x-ray image of the carabid beetle *Platynus decentis* head and thorax (top view), with largest thoracic respiratory tracts highlighted in red. Modified from [23]. (**b**) Photograph of insect-inspired microfluidic device (design M2). The red color represents the actuation network and green color represents the insect-inspired fluid network (highlighted in Figure 1(a)). (**c**) Time series images (1–4) of directional tracheal compression in the horned passalus beetle, *Odontotaenius disjunctus*. Modified from [29]. Collapse propagates from lower left of image (red point pair) to upper right (yellow point pair). (**d**) Synchrotron x-ray image of the largest thoracic tracheae, fully inflated, in the carabid beetle, *Pterostichus stygicus*, from [24]. (**e**) Synchrotron x-ray image of the thoracic tracheae, now fully compressed, with two discrete collapse locations indicated.

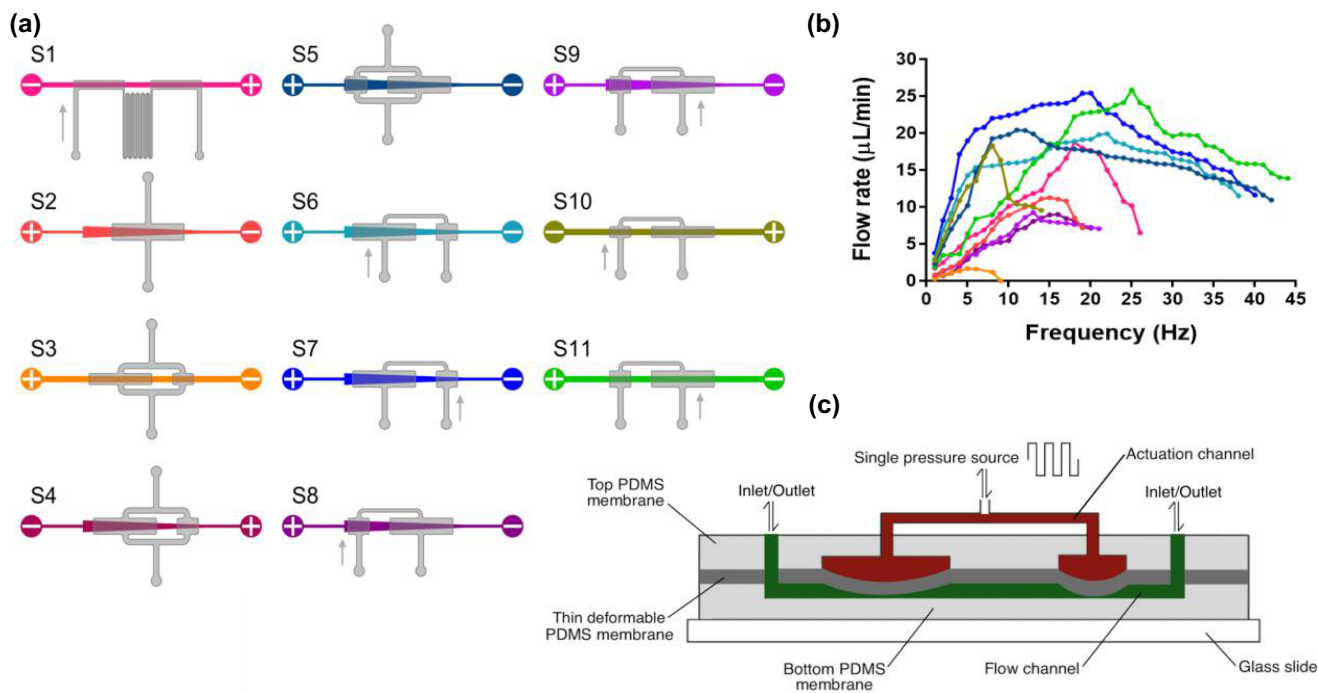

**Figure 2 Single channel devices.** (**a**) Schematics of the eleven single channel devices. (**b**) Flow rate versus frequency for all single-channel devices except S4. (Curves color coded to match device schematics in (a); $\Delta p$ = 6.5 $\pm$ 1.5 psi for devices S1–3 and S5–11). (**c**) Schematic (side view, not to scale) of three-layer polydimethylsiloxane (PDMS) device. A single pressure source provides periodic pressurization and evacuation of the actuation channels (maroon), deflecting a thin PDMS membrane (dark gray) and generating flow through the insect-inspired network (green). Channel depth is 80 $\mu$m for all devices, width varies from 200–1000 $\mu$m.

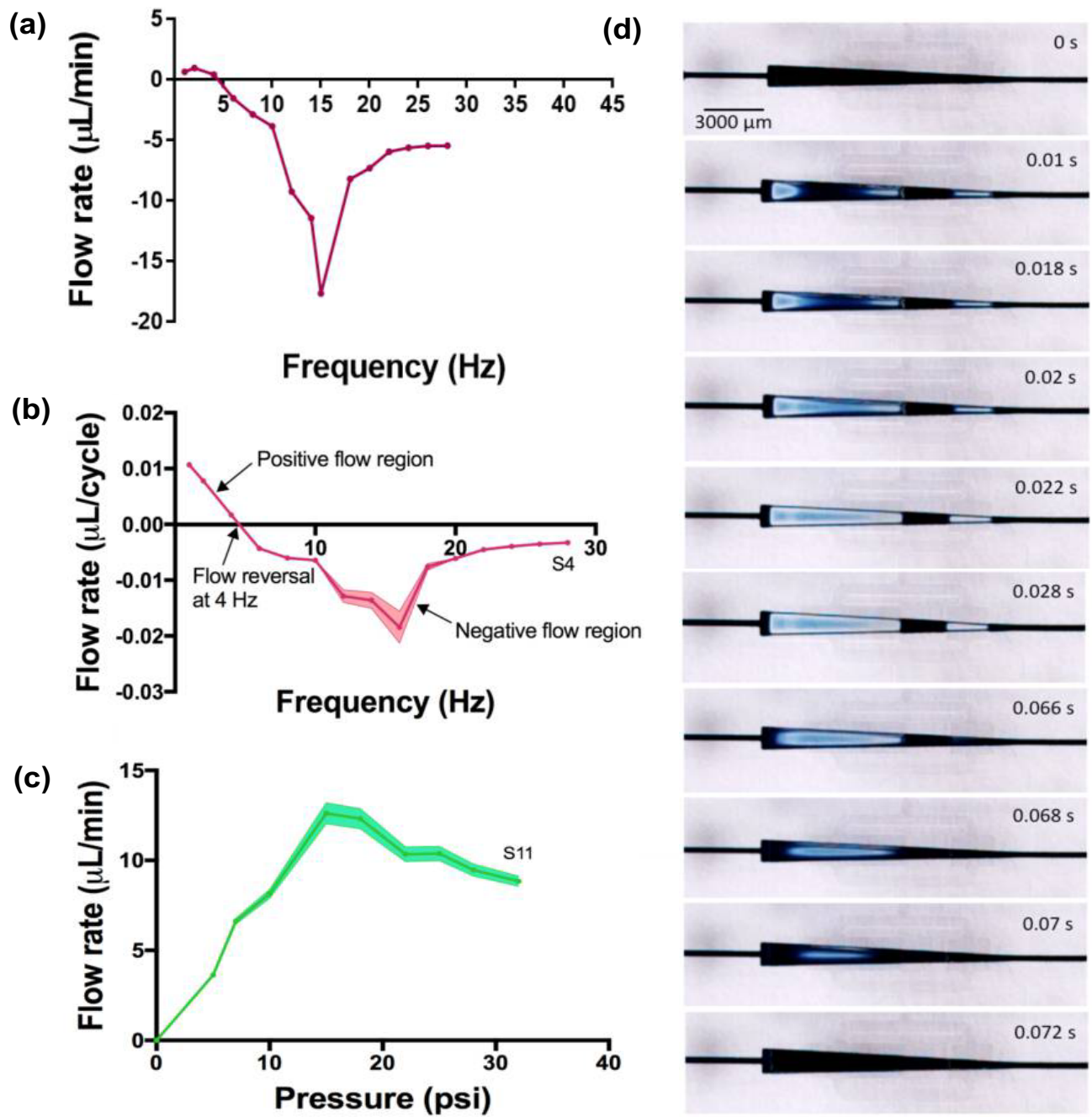

**Figure 3 Performance of devices S4 and S11**

(**a**) Flow rate versus *f* for device S4 ($\Delta p$ = 10.0 $\pm$ 1.0 psi). The flow in device S4 reverses direction above a critical frequency of approximately 4 Hz. (**b**) Flow rate per cycle versus *f* for device S4 ($\Delta p$ = 10.0 $\pm$ 1.0 psi). (**c**) Flow rate versus ($\Delta p$ for device S11 at *f* = 4 Hz. Shading ((e) and (f)) represents the error due to the variance of the data. (**d**) Top view of device S4 over a complete collapse cycle at *f* = 7.81 Hz.

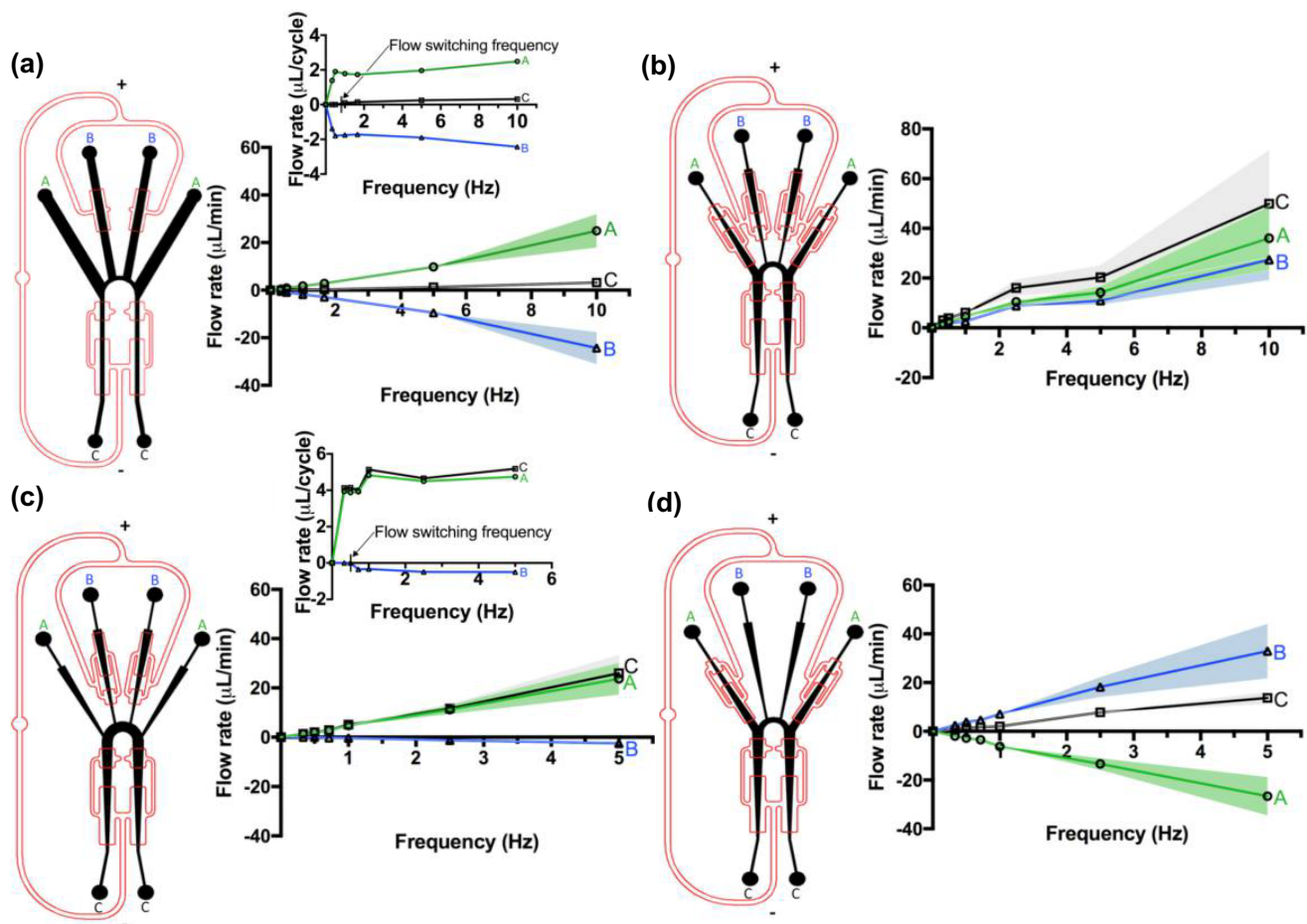

**Figure 4 Multichannel devices can switch flow in a branch on or off.** Schematics and flow rate data for four multichannel microfluidic devices. All devices are left-right symmetric. Positive flow is in the "+" direction. Shading represents the uncertainty in the data due to measurement error, which increases with flow rate. (**a**) Frequency-dependent channel switching in device M1. Inset shows the flow rate per cycle versus *f*. The switching behavior is seen more clearly on these axes. (**b**) Device M2 produces positive flow through all three channels for every *f* tested. (**c**) Frequency-dependent flow switching in device M3. (**d**) Device M4 produces negative flow through channel A and positive flow through channels B and C at every *f* tested. All devices were tested at $\Delta p$ = 14.0 $\pm$ 1.0 psi

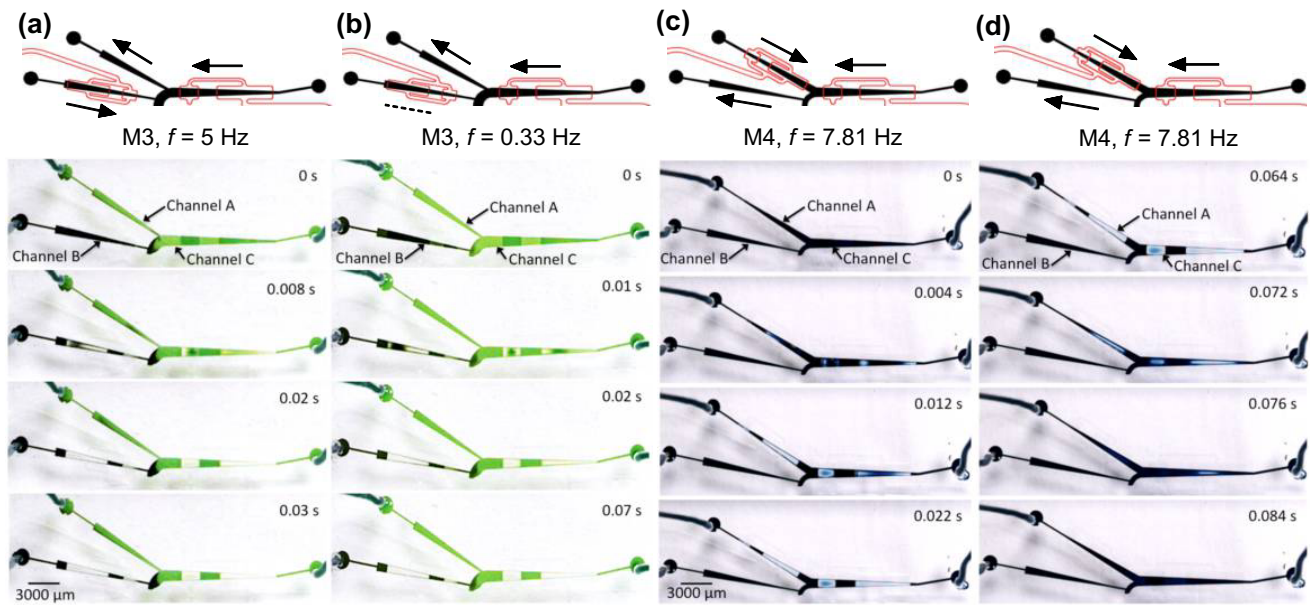

**Figure 5 Colorized tracheal network devices illustrate flow switching.** Right half of left-right symmetric networks shown. Flow direction convention same as in Fig. 3. (**a**) Device M3 actuated at high frequency (5 Hz). (**b**) Device M3 actuated at low frequency (0.33 Hz). (**c**) Device M4 actuated at 7.81 Hz, contraction part of cycle. Fluid was pumped from channels A and C into channel B. (**d**) Device M4 actuated at 7.81 Hz, re-expansion part of cycle. Fluid continued to be pumped from channels A and C into channel B. The actuation pressure for (a) and (b) was $\Delta p$ = 14.0 $\pm$ 1.0 psi while that for (c) and (d) was $\Delta p$ = 18.0$\pm$ 1.0 psi. A video demonstrating flow switching in device M3 can be found in Supplementary Materials.